\documentclass{article}

\usepackage{arxiv}

\usepackage[utf8]{inputenc} 
\usepackage[T1]{fontenc}    
\usepackage{hyperref}       
\usepackage{url}            
\usepackage{booktabs}       
\usepackage{amsfonts}       
\usepackage{nicefrac}       
\usepackage[disable=footnote]{microtype}      
\usepackage{lipsum}		
\usepackage{graphicx}
\usepackage{natbib}
\usepackage{doi}
\usepackage{amssymb,amsmath,amsthm,amsfonts}
\usepackage{setspace}
\usepackage{lineno}

\usepackage{xcolor}
\usepackage{soulpos}
\usepackage{ulem}
\ulposdef{\hlst}{%
    \rlap{\textcolor{yellow}{\rule[-.75ex]{\ulwidth}{2.5ex}}}%
    \rule[.45ex]{\ulwidth}{.1ex}%
}

\long\def\comment#1{}

\title{Detection of moving objects using self-motion constraints on optic flow}

\author{Hope Lutwak \\
	Center for Neural Science\\
	New York University\\
	New York, NY USA\\
	\texttt{hlutwak@nyu.edu} \\
	\And
	Bas Rokers \\
	Psychology, and Center for Brain and Health\\
	New York University Abu Dhabi\\
	Abu Dhabi, United Arab Emirates\\
	\texttt{rokers@nyu.edu} \\
	\And
	Eero P. Simoncelli \\
	Center For Neural Science, and \\
	Courant Inst. for Mathematical Sciences\\
	New York University, New York, NY USA \\
	Flatiron Institute, Simons Foundation  \\
	\texttt{eero.simoncelli@nyu.edu}
}

\date{}

\hypersetup{
pdftitle={A template for the arxiv style},
pdfsubject={q-bio.NC, q-bio.QM},
pdfauthor={David S.~Hippocampus, Elias D.~Striatum},
pdfkeywords={First keyword, Second keyword, More},
}

\begin{document}
\maketitle

\begin{abstract}
As we move through the world, the pattern of light projected on our eyes is complex and dynamic, yet we are still able to distinguish between moving and stationary objects. 
We propose that humans accomplish this by exploiting constraints that self-motion imposes on retinal velocities. When an eye translates and rotates in a stationary 3D scene, the velocity at each retinal location is constrained to a line segment in the 2D space of retinal velocities. The slope and intercept of this segment is determined by the eye’s translation and rotation, and the position along the segment is determined by local scene depth. 
Since all possible velocities arising from a stationary scene must lie on this segment, velocities that are not must correspond to objects moving within the scene. 
We hypothesize that humans make use of these constraints by using deviations of local velocity from these constraint lines to detect moving objects. To test this, we used a virtual reality headset to present rich wide-field stimuli, simulating the visual experience of translating forward in several virtual environments with varied precision of depth information.
Participants had to determine if a cued object moved relative to the scene. Consistent with the hypothesis, we found that performance depended on the deviation of the object velocity from the constraint segment, rather than a difference between retinal velocities of the object and its local surround. We also found that the endpoints of the constraint segment reflected the precision of depth information available in the different virtual environments. 
\end{abstract}


\section{Introduction}

When we move through the world, the spatial pattern of light arriving at our eyes moves accordingly, resulting in a pattern of velocities on the retina. This field of 2D retinal velocities, generally known as \textit{optic flow}, is a rich source of information that we use to understand and interact with the 3D world \citep{gibson1950perception}. During self-motion, retinal flow arises from both stationary and moving objects. Prior work has provided evidence that observers can reliably detect moving objects during self-motion \citep{rushton2007pop}, although estimates of object velocity may be biased  \citep{warren2009optic}. Here, we introduce a computational constraint that can be used to distinguish moving from stationary objects under general conditions, and conduct experiments to probe its influence in human observers.
 
\begin{figure}
  \centering
  \includegraphics[width=1\linewidth]{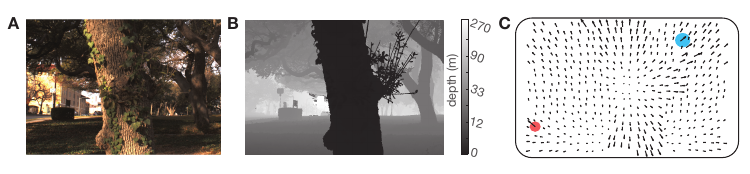}
  \caption{\textbf{Simulated optic flow field for an observer moving in a rigid environment.} (A) Photograph of an outdoor scene, taken from \citep{burge2011optimal}. (B) Depth map of the image in panel A (intensity proportional to log of depth) as measured  with LiDAR. (C) Calculated optic flow vectors for an observer moving forward and upward. Two highlighted vectors correspond to an artificially inserted independently moving object (red), and a location in the scene that is adjacent to a large depth discontinuity. Both highlighted velocities differ substantially (in both speed and ) from their average surrounding velocity.}
  \label{fig:flow objects}
\end{figure}

Figure \ref{fig:flow objects} shows a simulation of a complex optic flow pattern resulting from a translating, fixating observer in a mostly rigid 3D scene. The simulated observer moves forward and upward, and the foreground flow (arising from points on the tree) is dominated by radially outward velocities. Velocities of points near fixation (center of the tree trunk) are necessarily small. Velocities of points in the background are mostly small and upward, due to the rotation of the fixating eye. Given the general smoothness of the flow field, an intuitive strategy for detecting moving objects is to look for velocities that differ substantially from those of surrounding locations.  In this example, we simulate a moving object in the lower left corner of the scene (e.g., a bird taking flight from the ground), which produces an outlier in the flow field (velocity vector highlighted in red, Figure \ref{fig:flow objects}C). This would be correctly detected as a moving object due to the difference in speed and direction relative to surrounding velocities. However, this simple, local strategy will also detect velocity discontinuities at occlusion boundaries of fixed objects, such as the highlighted tree branch (blue). Velocities of points on the branch also differ in both speed and direction from surrounding locations, because many of the surround velocities are dominated by the rotation of the eye (they result from scenery that is further away), and the velocity of the branch in the foreground is dominated by the translation forward. \comment{compared to many of the surround velocities because those result from scenery that is further away, and the flow is dominated by the rotation of the eye. }
How can we distinguish variations in the flow field that arise from objects moving through a scene from those that arise from self-motion? 

We hypothesize that the visual system distinguishes these two situations by exploiting the constraints imposed on local velocities by self-motion in a stationary world. Specifically, the set of instantaneous velocities that could arise from self-motion within a stationary world all lie on a line segment within the space of 2D velocities \citep{longuet1980interpretation,heeger1988egomotion}. The attributes (slope, intercept) of this segment depend on the 3D motion of the eye through the world, while the position along the segment is determined solely by the distance to the point in the world. Since this segment contains all velocities consistent with motion through a rigid world, any observed velocity that does not lie on this segment must correspond to an independently moving object. To our knowledge the hypothesis that our ability to distinguish such independent motions should depend on the deviation of their projected velocities from the constraint segment has not been tested. Beyond providing a method to detect independently moving objects, constraint segment length can also reveal an observer's estimate of depth. Sections of the constraint segment correspond to ranges of depth in the world, so if an observer had a good estimate of depth, they could use a smaller segment of the constraint to determine which velocities must originate from moving objects. The interpretablility of the constraint segment makes it a powerful framework to test perceptual capabilities.

In this paper we examined whether observer performance in detecting moving objects can be explained by their reliance on the constraint by presenting a simulated translation through several rendered scenes. By manipulating a target object's 3D motion, which in turn varies its 2D retinal velocities, we evaluated whether distance to the constraint explains performance better than their similarity to the surround velocities. We also varied the availability of depth information in the rendered scenes and determine whether uncertainty in depth was consistent with an increase in the effective length of the constraint segments. We found evidence in support of both hypotheses.

\section{Methods}

\subsection{Optic flow for a fixating observer in a rigid world}
\begin{figure}
	\centering
	\includegraphics[width=.75\linewidth]{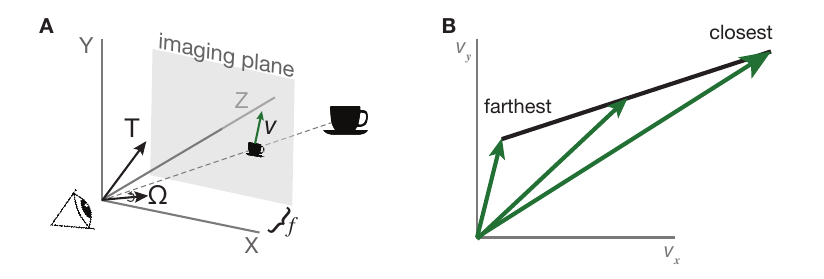}
	\caption{\textbf{Eye-centered projective geometry defines the depth constraint line on image velocity.} (A) Schematic of a single fixating observer undergoing 3D translation $\mathbf{T}$ and rotation $\mathbf{\Omega}$. A stationary object in the world (teacup) is projected onto an image plane (gray skewed rectangle) at distance \textit{f} from the eye. The velocity $\textbf{v}= (v_x, v_y)^\intercal$ of the projected object on the image plane is determined  by both the translation and rotation of the observer. If the object is moving independently in the world, that velocity also contributes to its velocity on the image plane. (B) The velocity at each image location depends on the distance (depth) of the corresponding 3D world location, and the set of possibilities lie along a single \textit{depth constraint line segment} (solid black line) in the 2D space of velocities.}
	\label{fig:image plane}
\end{figure}

In order to derive the velocities for a translating fixating observer, we start with the projective geometry of a pinhole camera, and use it to calculate the velocities across the visual field for any translation and rotation of the eye in a rigid world \citep{longuet1980interpretation,heeger1992subspace}. Points in the world are expressed as $(X,Y,Z)$, in eye-centered coordinates.  These are projected onto an image plane at coordinate location $(x,y)$, a distance $f$ away from the eye as depicted in Figure \ref{fig:image plane}A.

The image velocity $\textbf{v}$ of a point in the rigid world that projects to image position $(x,y)$ can be written as:
\begin{equation} \label{eq:opticflow}
    \textbf{v}(x,y) = \frac{1}{Z(x,y)}   \textbf{A}(x,y) \textbf{T} + \textbf{B}(x,y) \boldsymbol{\Omega} , 
\end{equation}
where $\textbf{T} = (T_X, T_Y, T_Z)^\intercal$ and $\boldsymbol{\Omega} =(\Omega_X, \Omega_Y, \Omega_Z)^\intercal$ are vectors expressing the translation and rotation components of self-motion in eye-centered coordinates. The matrices \textbf{A}$(x,y)$ and \textbf{B}$(x,y)$ are dependent on $f$ and $(x,y)$:
\begin{equation} \label{eq:AB}
\textbf{A}(x,y) = \begin{bmatrix}
-f & 0 & x \\
0 & -f & y
\end{bmatrix}\quad \text{ and }\quad
\textbf{B}(x,y) = \begin{bmatrix}
\frac{xy}{f} & -(f+\frac{x^2}{f}) & y\\
f+\frac{y^2}{f} & -\frac{xy}{f} & -x
\end{bmatrix} .
\end{equation}

Primates fixate their gaze on points in the world, and this natural behavior allows us to simplify the parameterization of the rigid-world velocity field in Equation \eqref{eq:opticflow}. During fixation, the velocity at the origin of the image plane is by definition $\textbf{v}(0,0) = \vec{0}$. Incorporating this constraint into Equation \eqref{eq:opticflow} and solving for elements of $\Omega$ gives
\begin{equation} \label{eq:omegas}
\Omega_X = \frac{1}{Z_0}T_Y \quad\text{ and }\quad \Omega_Y = -\frac{1}{Z_0}T_X ,
\end{equation}
where $Z_0 = Z(0,0)$, the depth at fixation. We assume the remaining (torsional) component of the eye rotation is zero (i.e., $\Omega_Z = 0$). Related derivations of flow during gaze fixation can also be found in previous publications  \citep{thomas1994spherical,calow2007local}. 
Under these conditions, the velocity field can be written solely as a function of $\textbf{T}$:
\begin{equation}\label{eq:flow_fixation}
  \textbf{v}(x,y) = 
    \frac{1}{Z_0}
    \left[
    \begin{array}{ccc}
    f+\frac{x^2}{f} & \frac{xy}{f} & 0 \\
    \frac{xy}{f} & f+\frac{y^2}{f} & 0
    \end{array}
    \right]
    \textbf{T}
    +
    \frac{1}{Z(x,y)}
    \left[
    \begin{array}{ccc}
    -f & 0 & x \\
    0 & -f & y
    \end{array}
    \right]\textbf{T} .
\end{equation}
%
%
If we assume the observer's translational velocity, ${\bf T}$, is known, the velocity at a given location $(x,y)$ is expressed as a sum of a fixed vector, and a second vector whose length is inversely proportional to depth. That is,  lies along a line, in the direction of the second vector. Since depth must lie in the interval $[f, \infty]$, inverse depth should lie in the interval $[0, 1/f]$, and thus the set of possible velocities lie within a line segment corresponding to this range of inverse depths (see Figure \ref{fig:image plane}B). Note that this constraint segment is fully determined by the translation vector (${\bf T}$), the position on the image plane $(x,y)$, and the range of allowable depths.



The depth constraint segment describes all the possible image plane velocities that can arise for a translating, fixating observer moving through a rigid world. Any velocities that do not lie on the constraint segment must therefore be due to object motion. We would like to know if humans make use of this constraint, by inferring their 3D self-motion from the global pattern of retinal velocities, and using deviations of local velocity from the corresponding constraint to detect objects in motion. To do this, we tested observers' ability to correctly distinguish independently moving objects from an otherwise stationary environment during simulated self-motion. We expected better performance when the 2D retinal velocities caused by the independently moving object are very far from their respective constraint segments. On the other hand, we expected worse performance when the distance to the constraint is small. We also expected that the length of the constraint segment assumed by a human observer would vary based on the amount of depth information provided by the stimuli.

\subsection{Stimulus Design}\label{exp2stimdesign}

\begin{figure}
	\centering
	\includegraphics[width=1\linewidth]{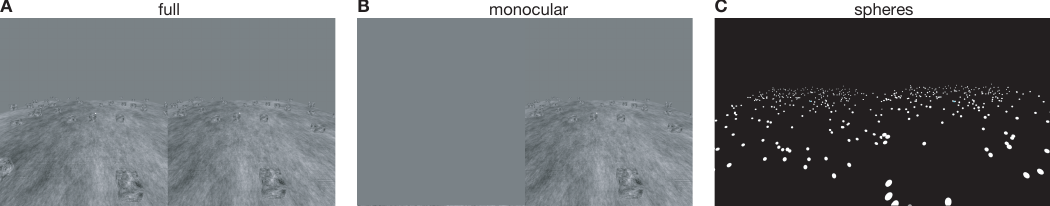}
	\caption[Binocular views of VR environments]{\textbf{Binocular views of VR environments} (A) Full stimulus with 1/f textured ground plane and randomly placed cubes. (B) Monocular stimulus with textured ground plane and cubes, shown to the right eye.  (C) Spheres stimulus with 0.025 m diameter spheres scattered on the ground plane as well as one sphere placed randomly in the same place as each cube in the full and monocular conditions.}
	\label{fig:exp2vrview}
\end{figure}

We simulated an observer translating forward at 1 m/s in three different virtual reality (VR) environments. The \textbf{full condition} (Figure \ref{fig:exp2vrview}A), consisted of a ground plane (6 m x 6 m) appearing 1 m below the observer and rendered with a 1/f noise pattern, against an isoluminant gray background. The full condition included 50 randomly scattered cubes (each with 0.15 m side length), rendered with 1/f noise textures on each side. Cubes floated above the ground plane at random height sampled from a uniform distribution ranging from 0 to 0.15 m. The \textbf{monocular condition} (Figure \ref{fig:exp2vrview}B) was the same as the full stimulus, but virtual objects (ground plane and cubes) were only rendered in the right eye. The left eye viewed an isoluminant gray background (same as that of the background in the full condition). Lastly the \textbf{spheres condition} consisted of 150 white spheres (0.025 m diameter) scattered randomly on or above an invisible (black) ground plane (again, 6 m x 6 m). Each sphere was generated in the same way as the cubes were in the full condition. In all conditions, observers were asked to maintain fixation on a sphere lying in the center of the ground plane (0.025 m diameter, which was rendered in white in the full and monocular conditions, and cyan in the spheres condition (to distinguish it from the other spheres). 


\begin{figure}
	\centering
	\includegraphics[width=1\linewidth]{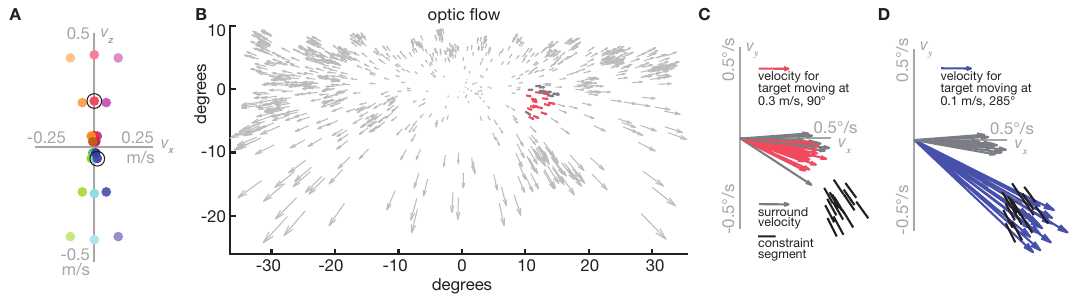}
	\caption[Target object speeds and directions tested]{\textbf{Target object speeds and directions tested.} (A) Bird's eye view of target motions in the X-Z plane. Z+ are directions away from the viewer, while Z- are directions towards the observer. (B) Optic flow for a translating fixating observer within environment consisting of textured ground plane, and cubes, with a target cube moving independently $0.3$ m/s away from the observer ($90$°, outlined red point in  panel {A}. (B) Velocities of points on the target cube (red), and surround velocities within a $3.5$° radius (dark gray). Velocity vectors are drawn scaled up by a factor of 20 for visualization. (C) Moving target (red) and surround (gray) velocity vectors, redrawn in a common coordinate system, along with constraint segments associated with the target locations. For this target motion, the target velocities are similar to those of the surround, but are far from the constraint segments. (D) Target and surround velocities for a target moving at $0.1$ m/s largely toward the observer ($285$°, outlined blue point in panel {A}. In this condition, target velocities lie near the constraint segments, but far from the surround velocities.}
	\label{fig:exp2conditions}
\end{figure}

On each trial a target cube (or sphere) appeared 0.5 m to the left or right randomly, 3 m away, and 0.15 m above the ground plane. A mirrored stationary cube (or sphere, in the spheres stimulus) was placed on the opposite side at the same 3 m distance, and 0.15 m above the ground plane. The target cube moved in one of 24 selected directions in the X-Z plane (Figure \ref{fig:exp2conditions}A): speeds $[0.05, .1, .3, .5]$ m/s and directions $[75, 90, 105, 255, 270, 285]$°. These target motions were selected to yield 2D retinal velocities that would best distinguish strategies of comparison to the constraint segment versus comparison to the velocities of the surround. To see this, we first calculated the optic flow for each scene and stimulus condition (as described in Section \ref{exp2analysis}), and then compiled the target and surround velocities from the flow field as shown in Figure \ref{fig:exp2conditions}B. An example of a target motion that results in retinal velocities that heavily overlap with the surround velocities is shown in Figure \ref{fig:exp2conditions}C. The red vectors depict a sample of the 2D retinal velocities of the moving target (at $0.3$ m/s, away from the observer ($90$°)) combined with the observer translating forward and rotating their eyes to maintain fixation on the provided fixation point. Because of the target's motion and its placement above the ground plane, the surround velocities, shown in gray, are very similar. In this case if observers only compared the target velocities with the surround velocities, where there is little difference, they would not be able to detect this moving object. However if observers had access to the constraint segment (shown as black segments, corresponding to each sampled target velocity), they would easily be able to detect the moving object because the velocities do not lie on the constraint segment. In the same vein, in Figure \ref{fig:exp2conditions}D, the blue vectors are the 2D retinal velocities for a target object moving at 0.1 m/s 285°, and the same surround velocities are depicted in gray. An observer who compares to the surround in this scenario can easily tell that the target object is moving, while an observer who relies on the constraint will find the task more difficult since the velocities are close to their respective constraint segments. 


\subsection{Experimental Procedure}\label{ch4experimentalprocedure}

On each trial, participants observed a moving stimulus for 0.5 s, while tracking a fixation point in the center of the ground plane. Participants were seated and used a chin rest to stabilize their heads. Simulated observer motion was forward at 1 m/s, parallel to the ground plane, and participants had to rotate their eyes to maintain fixation on the provided point on the ground plane. Over the 0.5 s stimulus duration, the target object moved with velocity randomly selected from 24 motions(see Section \ref{exp2stimdesign}). Each target motion was repeated 40 times per environment over the course of two sessions (total of 960 trials per environment and participant). At the end of the stimulus presentation, the observer view was reset, and the objects disappeared. At the start of a new trial the fixation point and the ground plane appeared (or fixation only for the spheres stimulus). Observers initiated the start of a new trial with a key press once they had attained fixation. 

We collected data over two sessions, with each session consisting of one block for each environment. Presentation order of the environments was counterbalanced. For the first session, participants performed 10 practice trials at the start of each block where the target object moved at 1 m/s in one of the 6 directions. For the second session, participants performed 5 practice trials to start. 

\subsection{Display, Eye tracking, and Software}

Stimuli were presented on a stereoscopic virtual reality headset (Oculus CV1). The headset has two Pentile OLED displays each with a refresh rate of 90 Hz and 1080x1200 pixels per eye. The horizontal and vertical FOV is 90$^\circ$. Before the first session we measured the participant's interpupillary distance (IPD) using an Essilor X81705 Pupillometer and set the headset according to the individual's IPD. 

We captured participants' eye movements using a Pupil Labs Neon device, which was run remotely with an Android smartphone (OnePlus 10 Pro 5G) with the Companion Device Pupil Labs custom software. The Pupil Labs Neon has two IR eye cameras, that capture at 200Hz at 192x192 pixel resolution. The Neon outputs gaze vectors relative to the scene camera based on the IR camera images and their deep-learning based pipeline NeonNet, resulting in a stated 1.8° median accuracy at 1.3 m depth across a 60° x 35° field of view \citep{baumann2023neon}. Eye tracking data and response data were aligned by sending events through Matlab to the eye tracker. Eye movement data was synced with behavioral data based on world clocks. Participants wore the Pupil Labs Neon glasses underneath the Oculus CV1. The chin rest helped ensure a participant's head was stable relative to the headset and eye tracking device during each session.


Software was written in Matlab using the Psychophysics Toolbox and OpenGL using an ACER laptop running Windows 10. Analysis software was written in Matlab using the Psychophysics Toolbox144 \citep{brainard1997psychophysics,kleiner2007}

\subsection{Subjects}

All 10 (5 female, 5 male) naive observers had normal or corrected-to-normal vision. The experiment was conducted in accordance with the Declaration of Helsinki and all procedures were approved by the New York University Ethics Committee on Activities Involving Human Subjects.

\subsection{Analysis}\label{exp2analysis}

Prior to analysis we removed participants and trials when horizontal eye movements deviated substantially from the trajectory specified by the fixation point. To do this, we first applied a fourth order Butterworth filter with a 0.1 to 60 Hz bandpass to remove high frequency noise as well as drift due to headset shifts over the course of a block. Filtered eye tracking data is shown in the Supplement, Section \ref{exp2supplement}, Figure \ref{fig:exp2eyetracking}A.  We removed trials in which horizontal gaze position deviated from the average starting or ending horizontal gaze direction by more than 1.5°. This occurred primarily when a subject made a horizontal saccade towards the moving target, which appeared about 13° from the center of gaze at the beginning of the trial. Data from two participants were removed since their eye tracking data did not fulfill our criterion. Across the remaining eight participants, 30\% of the full, 25\% of the monocular, and 16\% of the spheres trials were removed.


To measure the difference in target 2D velocities compared to the target's surround or the relevant constraint segments, we simulated optic flow vectors based on the scene structure of the three different VR environments. We placed 1000 points on a 6 m x 6 m ground plane, 1 m below the simulated observer. We sampled 15 points in the region occupied by the cubes, each cloud of points could be within a cube with side length of 0.15 m, same as in the virtual environment. For the spheres stimulus we did the same, but instead of point clouds, we just sampled one point at the location of each sphere. Then we simulated the observers translation forward through the environment at 1 m/s, sampled at 90 Hz (same as the refresh rate on the VR headset). For the main results, we rotated this simulated eye based on an eye rotation that would occur if the participant kept fixation on the provided point. Based on this eye translation and rotation, we calculated where the points would project onto an image plane and converted the positions to visual angle. To calculate the optic flow, we took the difference in positions between the frames. The resulting optic flow from the first two frames is shown in Figure \ref{fig:exp2conditions}B. For the supplementary results, we rotated the simulated eye based on the actual vertical component of the eye movements captured by the eye tracking system. 

\begin{figure}
	\centering
	\includegraphics[width=.8\linewidth]{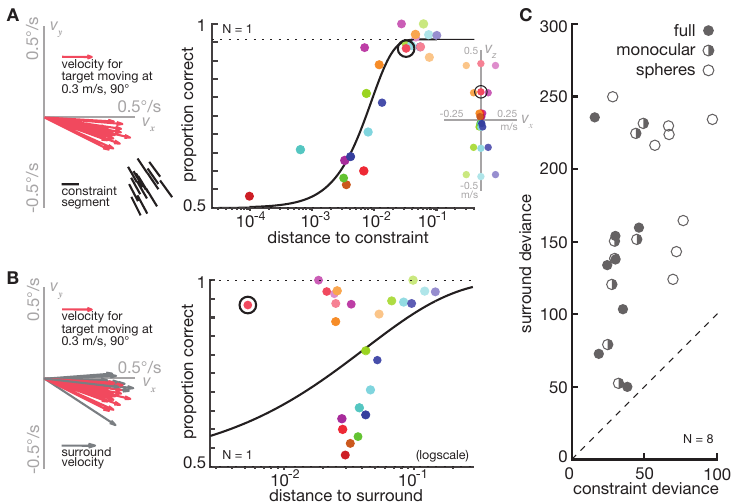}
 	\caption[Distance to the constraint versus distance to the surround]{\textbf{Distance to the constraint versus distance to the surround.} (A) Left: Velocities of target points (red) for a target moving $0.3$ m/s away from the observer, along with the constraint segments for each velocity (black). Right: The psychometric fit, with data from all target motion conditions (inset) arranged according to distance to the constraint. The circled data point corresponds to the target velocities plotted on the left. (B) Left: Same target velocities as panel A (red),  but now shown with corresponding surround velocities (gray). Right: Psychometric fit based on calculating the distance to the surround.(C) Comparison of deviance of fits based on the distance to the constraint vs. distance to the surround for each subject and stimulus condition. For every comparison, the distance to the constraint has a much better fit (lower deviance) than the distance to the surround (higher deviance). 
    }
	\label{fig:results_surr}
\end{figure}

To measure the \textit{distance to the constraint}, we first calculated the constraint segment associated with each sampled point on the target object (e.g., black segments in Figure \ref{fig:results_surr}A). While the full depth constraint segment encompasses all possible velocities for a particular point in the visual field, we instead calculated a truncated constraint segment corresponding to a depth range with a factor increase/decrease of 1.01 (±0.02 m) for results discussed in Section \ref{exp2results1}. \comment{We allowed this depth range to vary for results in Section \ref{exp2results2}.} We then calculated the smallest distance of each sampled target velocity to its corresponding constraint segment, and averaged over those distances. To obtain a singular measurement for the duration of the stimulus, we also averaged over all frames. To measure \textit{distance to the surround}, we defined the surround as the velocities within a $3.5$° radius of the center of the target cube on each frame, as shown in Figure \ref{fig:exp2conditions}B. We then collected all the vectors together, re-centered at the origin (Figure \ref{fig:results_surr}). From these we took the average of the target velocities as well as an average of the surround velocities, and then took the Euclidean distance between each, across all frames to obtain the distance to the surround. Observer performance was then assessed as a function of distance to the constraint or distance to the surround (Figure \ref{fig:results_surr}). Finally, we fit a Weibull function to these data, computed with the Matlab toolbox psignifit \citep{schutt2016painfree}. We used deviance, a goodness-of-fit measurement based on the probability that the psychometric fit could have produced the data, to determine if the distance to the constraint or distance to the surround better described behavior. 

\section{Results}

\subsection{Distance to the depth constraint segment can accurately describe performance}\label{exp2results1}

Figure \ref{fig:results_surr} shows results for a single subject.  Qualitatively, we can see that analyzing the data in terms of distance to the constraint (Figure \ref{fig:results_surr}A) provides a better account of behavior than analyzing in terms of distance to the average of the surrounding velocities (Figure \ref{fig:results_surr}B). While plotting the data in terms of distance to the constraint results in a distinct sigmoidal curve, this is not the case when plotting the data in terms of distance to the surround. For example, when the target moved at $0.3$ m/s, away from the observer, the target 2D velocities are similar to the surround velocities (has a small distance to the constraint value), but performance is still high (nearly 90\% correct). Results for each participant can be seen in the Supplement (Section \ref{exp2supplement}).  We quantified the goodness-of-fit by the ``deviance'',  a monotonic transformation of the likelihood of the model parameters of the fitted curve. It ranges from 0 (no error) to infinity \citep{wichmann2001psychometric}. For this individual, the deviance for the distance to the constraint fit is $30$, while the deviance for the distance to the surround is $135$. For all eight subjects, across all three VR environments tested, the deviance was substantially larger when fitting the data in terms of distance to the constraint compared to distance to the surround (Figure \ref{fig:results_surr}C).






\subsection{Length of the constraint segment}\label{exp2results2}

In the previous section we established that performance varied with the distance between the 2D retinal velocities to the constraint segment according to a psychometric function. We also hypothesized that the subsection of the constraint line that observers used reflects their internal estimate of local scene depth. Specifically, we interpret the endpoints of the velocity constraint segment as reflecting the range of depths that the visual system deems ``plausible'' in explaining the visual input in terms of stationary objects. 
Put differently, the midpoint of the constraint segment can be interpreted as an indication of the depth estimate of a visual field location, and the length as an indication of the uncertainty regarding that estimate.
Thus, if observers have access to less precise depth information (due to structure of the scene, self-motion, or visual stimulus construction), we expect them to consider that a velocity could come from a stationary object located at a wider range of candidate depths, which corresponds to a longer velocity constraint subsection. For the stimuli in our experiments, we might also expect the range of depths to be biased toward an estimate that is closer to the depth of the ground plane. As a result, performance in distinguishing a moving and stationary target should be best explained in terms of distances to constraint segments that are both extended, and shifted toward the velocity associated with the ground plane.



We conducted three versions of the experiment with conditions that conveyed different amounts of scene depth information. The {\bf full stimulus} condition included full textured cubes above a textured ground plane, with depth cues arising from relative size, motion parallax, and binocular disparity. The {\bf monocular stimulus} condition presented the same right eye display as the full stimulus, and a uniform gray background in the left eye. Finally the {\bf sphere stimulus} degraded the monocular depth cues of the full stimulus by replacing the textured ground plane with small white spheres as well a single sphere centered within each cube. This scene was displayed stereoscopically. 

\begin{figure}
	\centering
	\includegraphics[width=1\linewidth]{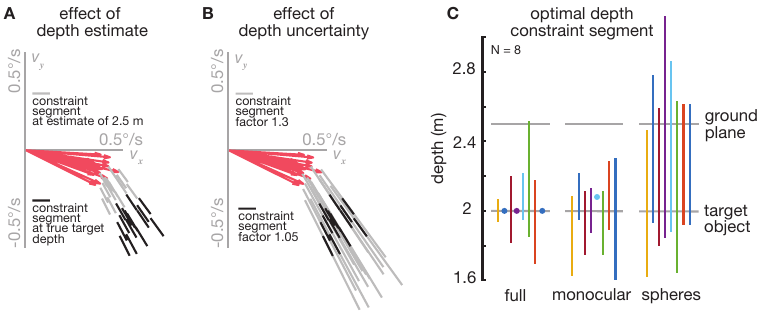}
 	\caption[Varying the mean depth and depth range]{\textbf{Varying the mean depth and depth range.} (A) Illustration of effect of mean depth estimate on velocity constraint segment.  Shifting the mean depth of the target further away leads to constraint segments that are closer to the target velocity (i.e., the depth-dependent velocity component of Eq.~\ref{eq:flow_fixation} is reduced). (B) Illustration of the effect of increasing the depth range from a factor of 1.05 to a factor of 1.3.  (C) 
    Optimal depth constraint segment for all 8 subjects (identified by color) in each of the three VR environments. For all subjects, the fitted depth range was broader and shifted further away for the spheres stimulus.}
	\label{fig:exp2results_depthrange}
\end{figure}

For each participant and 3D environment, we recalculated the distance to the constraint segment corresponding to different depth ranges. We varied the mean depth over a range of $[-0.4,0.4]$ m relative to the true object depth (2 m). An illustration of shifting the mean depth further, while keeping the depth range fixed, is shown in Figure \ref{fig:exp2results_depthrange}A. In this case, where the target is moving at $0.3$ m/s away from the observer, the constraint segments were closer to the target velocities so the distance to the constraint was smaller. We also varied the depth range, using multiplicative factors ranging from $1.001$ to $1.4$. Example constraint segments for a depth range of $1.3$ are shown in \ref{fig:exp2results_depthrange}B. Extending the depth range and shifting the mean depth can have the same effect on the distance to the constraint for one target motion (in this case a decrease), however note that for a different target motion this will not be the case such as the secondary example described in Section \ref{exp2stimdesign} (target moving at $0.1$ m/s, $285$°). Across all combinations of depth mean and range (11 samples of mean, 10 samples of range), we sought values that minimized the deviance of the data from the psychometric fit (as defined by \citep{schutt2016painfree}) per 3D scene per subject. An example of this procedure for one individual is shown in the Supplement (Section \ref{exp2supplement}). 

On average, we found that monocular and full conditions yielded similar depth ranges as seen in (Figure \ref{fig:exp2results_depthrange}C). This is perhaps expected, since binocular cues are weak at the presented target distance (2 m). Furthermore, previous results indicate that inexperienced users tend to ignore binocular information in virtual reality headsets \citep{fulvio2020cue}. Finally, previous work has also found that 3D motion sensitivity is reduced in binocular compared to monocular viewing \citep{cooper2016sensitivity}. In line with our hypothesis, we found that the depth estimate that best explained our data was further away, and the depth range was broader for the spheres stimulus than for both the full and monocular stimuli (Figure \ref{fig:exp2results_depthrange}C), consistent with the corresponding uncertainty of depth information in these conditions. Based on other reports, decreasing the monocular information in a scene affected observers abilities to both detect moving objects \citep{royden2016effect}, as well as estimate object motion \citep{warren2009monocular} in optic flow fields that from translations forward. Our results are consistent with these findings, and provide a potential explanation for the reduced performance. 



\section{Discussion}
While moving through an environment, humans can reliably detect independently moving objects, despite the complexity of retinal motion patterns that can arise under such conditions. In this paper, we provide evidence that observers accomplish this by exploiting regularities in these patterns.  Specifically, each velocity vector associated with a non-moving point in the world must lie on a particular constraint line segment within the space of retinal velocities. We hypothesized, and provided experimental evidence, that human performance in detecting an independently moving object while moving in an otherwise rigid world varied monotonically with distance between its velocities and these line segments.
\comment{We hypothesized and provided evidence for observers' ability to take advantage of the constraints on retinal velocities based on the geometry of a fixating observer moving through the 3D world. We hypothesized that this constraint comes in the form of the depth constraint segment which describes all possible velocities in a rigid world for a given point in the visual field. An observer can then determine that any velocity which lies off the constraint segment must come from a moving object. We test this by examining if observers' performance on detecting a moving object can be described by the distance of the retinal velocities of the moving object to the respective constraint segments. We saw that the distance to the constraint could capture performance, especially over a local comparison to the surround velocities. 
}
We also found that under conditions of weaker depth cueing, observers behaved as if they were utilizing longer constraint segments, centered at depths biased toward that of the ground plane.

Previous work has shown that human ability to estimate the motion of independently moving objects can be biased by the radial flow expected for forward translational observer motion \citep{warren2008flowparsing, warren2009optic,warren2009monocular,rushton2018primary}. Based on a formulation dubbed \textit{flow parsing}, Warren and Rushton suggest that we are able to subtract out the optic flow that originates during forward self-motion in order estimate the velocity of moving objects. They also showed evidence for flow parsing when detecting object motion, even in the case when the surround velocities and moving target velocities are similar \citep{rushton2007pop}. Our framework provides a generalization of this hypothesis, operating on arbitrary translational observer motion, and allowing for uncertainties in depth. In particular, it offers an explanation of how large discontinuities in flow solely due to self-motion (e.g., occlusion boundaries of objects) can be distinguished from those arising from independently moving objects.




Our hypothesis relies on a sequence of three assumed computations:  Observers must (1) accurately estimate their translational motion, ${\mathbf T}$; (2) use this estimate to posit a velocity constraint segment associated with each retinal location; (3) compare local estimates of  retinal velocity (optic flow) with their corresponding constraint lines.
We believe that the wide-field display and dense-texturing of our stimuli, which produce a compelling subjective sense of 3D motion, played an important role in supporting the first of these assumed computations. Initial experiments performed on a conventional display, with a narrow field of view and isolated point motion stimuli, produced data that were inconclusive with respect to the hypothesis \citep{Lutwak2024thesis}. The proper temporal evolution of our stimuli may also play a role, consistent with recent results showing that realistic temporal evolution of motion improves heading estimation \citep{burlingham2020heading}. 

The velocity constraint segments in our formulation must lie along lines whose orientation and position is determined by the observers translational motion, but their endpoints are determined by the range of permissible depths.  We hypothesized, and found preliminary evidence, that these endpoints are not fixed but correspond to the observers' uncertainty about depth at each location. For the spheres stimuli, data were best explained by a depth range that was both broader than that for the other two stimulus conditions, and biased toward that of the ground plane (Fig. \ref{fig:exp2results_depthrange}C).
Consistent with this, previous studies found that inclusion of monocular depth cues, such as displaying a thin line connecting otherwise floating stimuli to the ground plane, resulted in lower thresholds for detecting moving objects \citep{royden2016effect, warren2009monocular}. A more systematic examination of the effects of monocular depth cues (or their absence) is warranted.

As an alternative to our hypothesis, we considered a simpler explanation based on comparison of retinal velocities of an object to those of the surrounding background.  Our experimental conditions included cases for which this strategy is effective (e.g., Fig. \ref{fig:exp2conditions}D) and others for which it is not (e.g., Fig. \ref{fig:exp2conditions}C).
We calculated the ``distance to the surround'' as the Euclidean distance between the average velocity over a randomly selected set of locations within the target, and the average velocity over a randomly selected set of surrounding locations. For simplicity, we chose surround locations within a 3.5° radius of the center of the object. We believe that allowing this radius to vary, or substituting a continuous but tapering weighting function (e.g., a Gaussian window), would not substantially alter the average surround velocities used in our analysis. A more sophisticated comparison could make use of differences in the velocity distribution other than the averages (e.g., earth-mover's distance between Gaussian distributions with the mean and covariance of the samples, or the Mahalanobis distance to measure saliency in terms of standard deviations from the mean surround \citep{rosenholtz1999simple}).   

Our subjects were head-fixed, and were asked to visually track a sphere lying on the ground plane during each 0.5s trial.
We sought to ensure consistent fixation by excluding trials with large horizontal eye movements, which occurred when subjects broke fixation to look at the target location. We had some concern that deviations (e.g., drift) in vertical fixation could  affect components of retinal velocity most relevant for our hypothesis, so re-analyzed the data with the measured eye movements included (see Supplement \ref{exp2supplement}). Under these conditions, we obtained the same result: distance to constraint captured performance better than distance to the surround. 
Further analysis of trials that were removed due to target saccades could potentially address questions related to foveated object motion detection \citep{brenner1991judging}.  An experiment in which fixation is not constrained, perhaps even with freely moving subjects
choosing their own environmental fixation targets (e.g., \citep{matthis2022retinal}), is substantially more difficult to design and control, but would offer a more natural experience, and a more diverse set of challenges for our hypothesis.

\comment{
One might also consider a model that fuses comparisons to both the surrounding velocities as well as the constraint segment. For example, in environments where an observer has minimal knowledge of depth information, or where heading estimates are more unreliable, one might expect observers to be more reliant on local velocity comparisons (e.g., \cite{Lutwak2024thesis}, chapter 3). 
That being said, when surround velocities are sampled close to a non-moving target in the visual field, those velocities should lie near the constraint segment, possibly at different depths.  In the case of the spheres condition, we found that the depth estimate that best explained our results lay closer to the ground plane. Thus measuring the distance to the constraint became more similar to our computation of distance to the surround under those conditions. 
}

The stimuli for our experiments simulated translation through a simplified scene anchored by a flat ground plane. Previous research has shown that the computational problem of depth-from-motion can benefit from parameterization relative to that of a reference plane, using a so-called ``plane plus parallax'' computation \citep{Kumar1994,Sawhney1994,irani2002direct}.
One estimates observer motion and a planar depth map that best explains the observed optic flow, and then interprets deviations in flow as arising from deviations in depth (i.e., parallax).
In the context of our experiments, our hypothesis offers a natural extension of this concept to human
detection of independently moving objects: deviations from planar flow are interpreted as arising from {\em either}
deviations in depth (parallax) or deviations in 3D motion (independently moving object).  It is of interest to examine whether and how these formulations extend to more complex 3D environments. 

We've defined a mathematical framework for the detection of objects that move within a stationary world, but it is of interest to consider how such computations might be achieved or approximated in the brain.  The dorsal stream of the primate visual system is generally believed to support the extraction and representation of motion and depth information \citep{Ungerleider83}.  Neurons in area MT are known to be selective for local retinal velocity \citep{Maunsell83,Movshon86,Bradley08}, whereas neurons in area MST are selective for patterns of flow \citep{Duffy91,Duffy95}, and have been proposed as a locus for the representation of heading direction \citep{Lappe96,Britten02}.
Recent work suggests that area MT neurons may provide signals useful for flow parsing
\citep{Peltier24}.
Another report finds that a subset of MT cells that are tuned incongruently for depth conveyed by disparity cues and motion cues, and could potentially provide a useful signal for detecting object motion \cite{kim2022neural}. 
\comment{As with the constraint segment of our hypothesis, there is a comparison of what should be true given a depth estimate (in this work disparity cues), and what the velocity actually is (motion parallax). If there is an incongruence, or in our terms a large distance to the constraint segment, then a moving object can be detected. They tested object motion detection where the retinal velocities for the moving object are very similar to stationary objects like the stimulus used in a perceptual experiment \cite{rushton2007pop}. In the same vein, congruent cells respond when the depth estimate and velocity signal align (on the constraint segment). } Our hypothesis provides a potential ecologically-driven interpretation for such incongruent selectivity, and can be used for a targeted experimental design for its validation or falsification.

Our hypothesis is constructed from the deterministic geometric relationships between optic flow, translating observer motion, and depth (Eq. \ref{eq:flow_fixation}), but could be generalized to a full ideal observer model, taking into account uncertainty in retinal velocity estimates, as well as uncertainty in the estimated translational observer motion, to detect non-rigid velocity outliers. Observers' ability to detect differences in speed and direction of flow is not uniform in 2D velocity space, or across the visual field. A first step would be to incorporate a  distribution for an estimated target velocity that adheres to Weber's Law for speed, and an angular component to describe directional sensitivity. There are also different sensitivities for motions along cardinal axes and relative reference frames \citep{ezzo2023asymmetries}. One could also incorporate more complex dependencies on the position within the visual field. Another source of error that we did not include arises from the estimate of heading, which has a rich history in the optic flow literature \citep{warren1988perception,warren1988direction}. It is known that (large) object motion can influence heading estimation \citep{warren1995perceiving, royden1996human, layton2016temporal}, and detecting object motion is generally assumed to rely on a reliable estimate of heading. More recent work has proposed methods of doing both at the same time \citep{scherff2024computational}. Testing observers' estimate of heading along with object motion detection would require a more sophisticated stimulus, but would open the door to more naturalistic environments and real participant self-motion. 

When we move through a stationary world, the retinal motion of image features obeys a strict set of mathematical constraints, and we have provided evidence that humans make specific use of these constraints to detect independently moving objects. The detection of independently moving objects is of critical importance to many animals, and our framework  thus offers an important step in elucidating how this capability is achieved.

\section{Acknowledgments}

The authors thank Mike Landy, Tony Movshon, and David Heeger for advice. Special thanks to Mike for experimental equipment as well. This work was supported by the NYUAD Center for Brain and Health, funded by Tamkeen under NYU Abu Dhabi Research Institute grant CG012 to BR,  a T32 grant from the NIH (EY007136-29) to NYU, and by the Center for Computational Neuroscience in the Flatiron Institute of the Simons Foundation. 

\appendix
\section{Supplement}\label{exp2supplement}

\subsection{Including vertical eyetracking in the analysis}

\begin{figure}[h]
	\centering
	\includegraphics[width=1\linewidth]{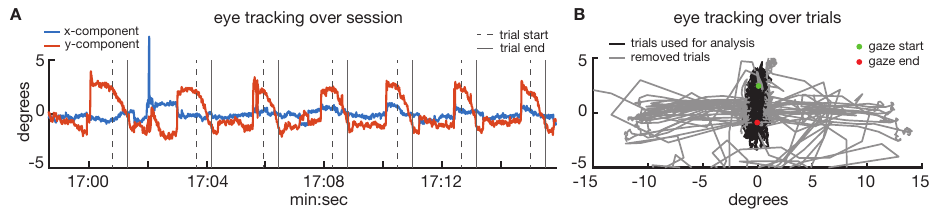}
	\caption[Eye tracking processing]{\textbf{Eye tracking processing procedure.} (A) 15 seconds of filtered eye tracking data taken from one subject. The x- and y-components of the gaze direction (°) are shown in blue and red, respectively, as a function of time. The dotted vertical lines mark the beginning of a trial, while the solid vertical lines mark the end of a trial. (B) Eye traces over all trials in a session from one subject. Traces that were removed are gray, while those within 1.5° of the average starting position (green dot) or ending position (red dot) are shown in black.}
	\label{fig:exp2eyetracking}
\end{figure}

For the main results, we removed trials that were more than 1.5° off from the average starting or ending horizontal gaze direction. Removed trials are shown in Figure \ref{fig:exp2eyetracking}B in gray.

\begin{figure}
	\centering
	\includegraphics[width=1\linewidth]{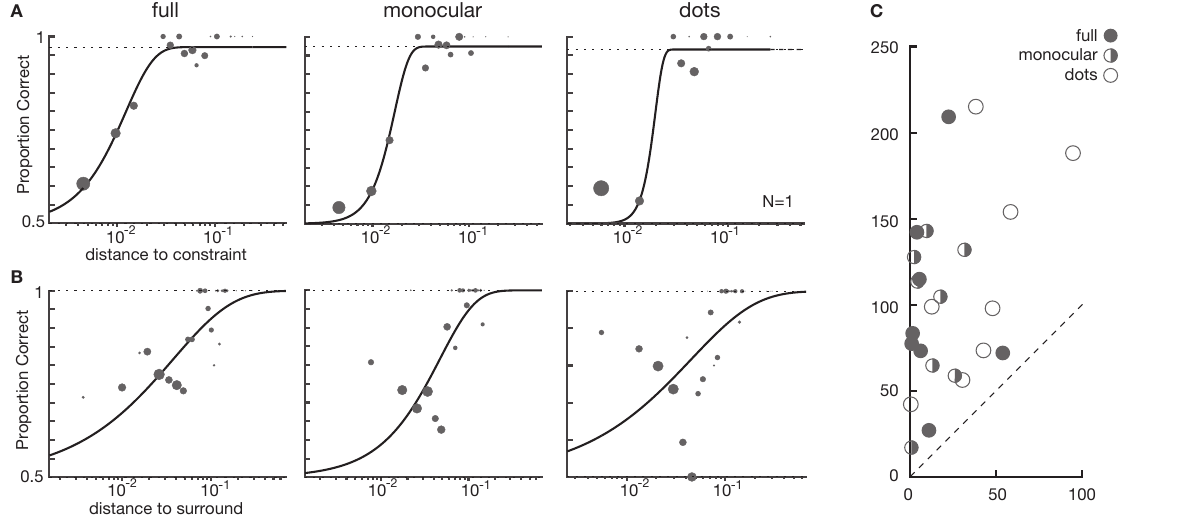}
 	\caption[Incorporating vertical eyetracking]{\textbf{Incorporating vertical eyetracking} Psychometric fits for each condition based on calculating the distance to the constraint (A) or distance to the surround (B) in each stimulus environment. (C) Deviance of fit based on the distance to the constraint compared to distance to the surround for each subject as well as each stimulus environment. For every comparison, the distance to the constraint has a better fit (lower deviance) than the distance to the surround (higher deviance). }
	\label{fig:exp2verticaleyetracking}
\end{figure}

All results were based on calculating ideal eye movements that result from the simulated translation forward, and perfect fixation on the given point 3 m away in depth on the ground plane. Subjects did not exactly adhere to this and made non-ideal eye movements. While trials with large horizontal eye movements were entirely removed, we used the behavioral data even if vertical eye movements strayed. Here we show the psychometric curves for one subject if we kept the actual vertical eye movements (Figure \ref{fig:exp2verticaleyetracking}A-B). Note that although our the formulation we used for the constraint segment is based on fixation (eye rotation $\Omega$ is dependent on $\bf{T}$), for any instant all velocities for a point still lie along a line when only $Z(x,y)$ is varied in Equation \eqref{eq:opticflow}.

Because eye movements vary from trial-to-trial, so does the distance to the constraint and surround. Data therefore are not color coded by target motion and are instead pooled solely by distance to the constraint (or surround). We can see that data generally are noisier, but the same result holds: the distance to the constraint describes performance better than distance to the surround. As seen in Figure \ref{fig:exp2verticaleyetracking}C, once again all data (except one point) have a worse fit (larger deviance) for when measuring distance to the surround compared to the constraint. More analysis would be required to investigate how eye movements may have contributed to making the task easier. For example, the lower values for the distance to the constraint are not actually as small as what they theoretically would be. Originally the smallest values of distance to the constraint were around $10^{-4}$, however with the veridical vertical eye movements, the lowest go down to the order of $10^{-2}$. 

\newpage
\subsection{Individual fits comparing distance to the constraint with distance to the surround}

The summary of the deviance for the psychometric fits across all subjects are shown in Figure \ref{fig:results_surr}C. Here we show all psychometric fits across all subjects, like the plots shown in Figure \ref{fig:results_surr}A-B. 

\begin{figure}
	\centering
	\includegraphics[width=1\linewidth]{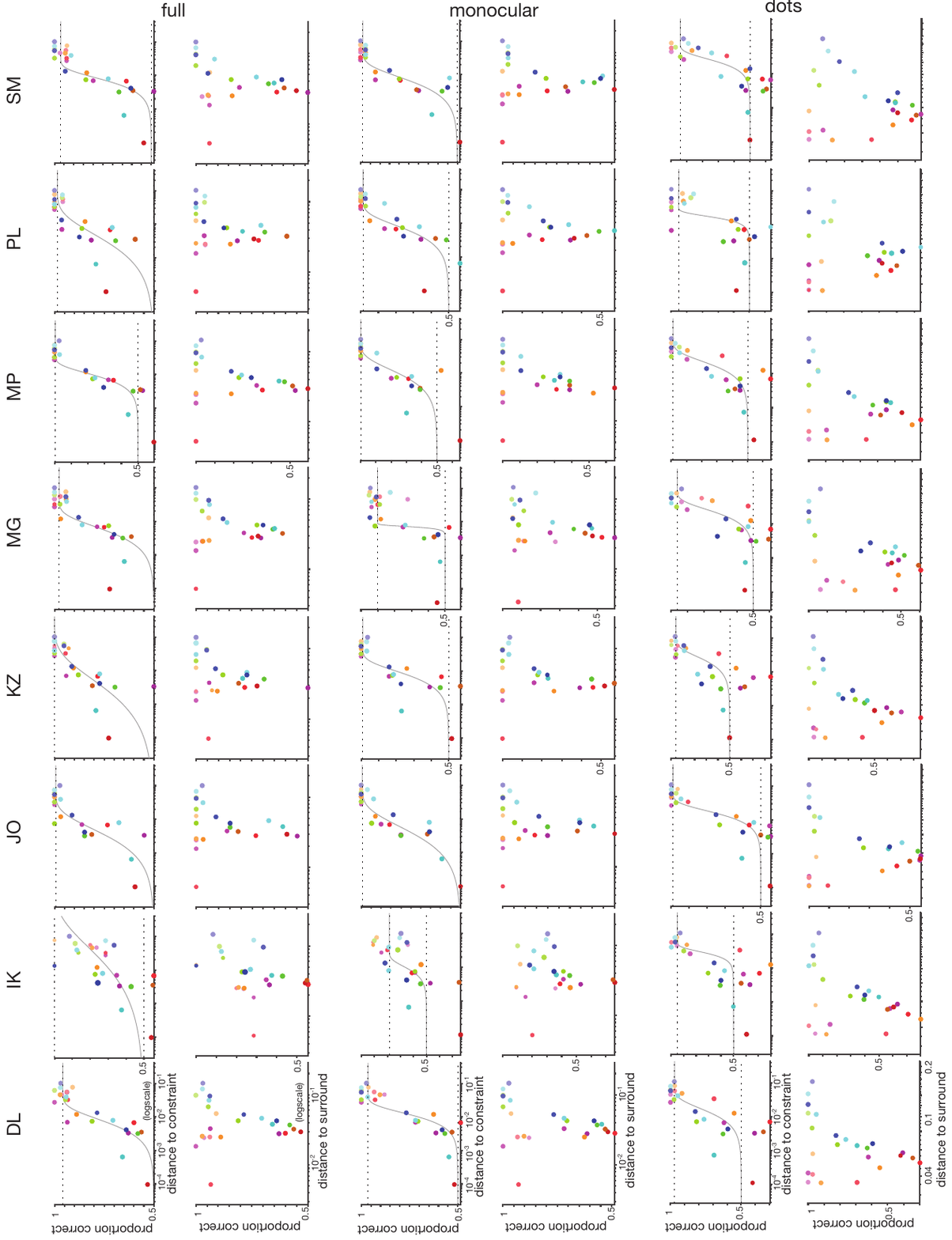}
 	\caption[Comparing psychometric fits across individuals]{\textbf{Comparing psychometric fits across individuals}}
	\label{fig:exp2individualfitsres1}
\end{figure}

\newpage
\subsection{Finding the optimal depth constraint segment}

\begin{figure}[h]
	\centering
	\includegraphics[width=1\linewidth]{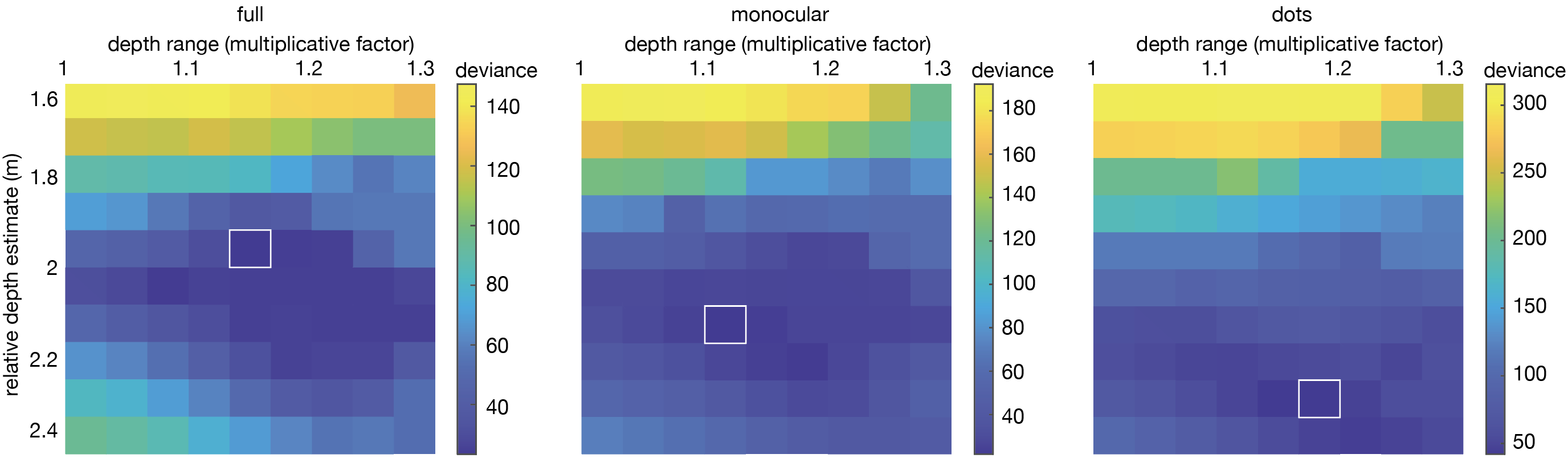}
 	\caption[Individual depth constraint optimization]{\textbf{Individual depth constraint optimization.} Deviance as a function of different depth estimates and depth ranges for one participant. For each 3D environment, we chose the values which minimized the deviance, highlighted in white.}
    
	\label{fig:exp2individual_depthoptimization}
\end{figure}

In Section \ref{exp2results2} we showed the results across participants for the optimal depth estimate and depth range. For each participant, we recalculated the distance to the constraint for a combination of different depth estimates and ranges. In Figure, \ref{fig:exp2individual_depthoptimization} we can see the deviance as a function of the different chosen depth estimates and ranges for one participant, for each 3D environment.

\bibliographystyle{apalike}
\bibliography{references}  

\newpage

\end{document}